\documentstyle[11pt,paspconf]{article}

\input{epsf}

\begin{document}

\title{Early Type galaxies in the Hierarchical Universe}

\author{C. M. Baugh, S. Cole, C. S. Frenk and A. J. Benson}
\affil{Department of Physics, Durham University, South Road, Durham, DH1 3LE, 
UK}

\author{C. G. Lacey}
\affil{Theoretical Astrophysics Center, Juliane Maries Vej 30, Copenhagen, 
Denmark}

\begin{abstract}
Any realistic theory of galaxy formation must be set in the context of 
a model for the formation of structure in the universe.
We describe a powerful approach -- {\it semi-analytic} modelling  
-- that combines a set of simple rules describing the gas processes 
involved in galaxy formation with a scheme to follow the 
hierarchical growth of dark matter haloes.
Surprisingly few free parameters are required to specify the model, 
and these are fixed with reference to a subset of local observational data.
The model produces the full star formation history of a galaxy, 
allowing a wide range of predictions to be made.
We review some of the successes of the models, namely the star formation 
history of the universe and the evolution of galaxy clustering, before 
focusing our attention on early-type galaxies.
We discuss the observational evidence against the 
classical picture in which early-type galaxies form at some 
arbitrarily high redshift in a single monolithic collapse and burst 
of star formation.
The alternative scenario in which spheroidal systems are formed by 
the merger of disk galaxies is outlined.
We review some of the predictions of this model, namely the colour-magnitude 
relation, the faint counts and the evolution of cluster membership.
\end{abstract}



\section{Introduction}
The classical picture of the formation of early type galaxies, 
in which the stars form in an intense burst at high redshift 
and then passively evolve to the present day, was worked out in the 
late 1960's and early 1970's (see for example Larson 1975 and 
references therein). 
Since this time, significant advances have been made in our understanding 
of the formation of structure in the universe. 
This is due to two reasons; the development of powerful theoretical 
techniques to follow the development of cosmological structures.
and the completion of increasingly large surveys of the local universe.

Several galaxy surveys have recently been completed that probe structure 
in the universe to a depth of $100-150h^{-1}$Mpc (e.g. Saunders etal. 1991; 
Loveday etal. 1992). The largest of these contains more than 20000 redshifts  
(Schectman etal. 1996). The next generation of redshift surveys, which will be 
completed early in the next millenium, will contain an order of 
magnitude more redshifts. 
Angular catalogues currently probe even larger volumes; 
the APM Survey contains two million galaxies and extends to more 
than $400 h^{-1}$Mpc (Maddox etal. 1996).
These surveys allow galaxy clustering to be measured on scales of tens 
of megaparsecs, which can reveal the nature of the primordial density 
fluctuations, subject to the uncertainty in the way that the galaxies 
trace out the underlying density field. 
Clustering studies at high redshift are also now becoming a reality, 
as exemplified by the measurement of strong clustering in the 
population of Lyman-break galaxies at $z=3$ (Adelberger etal. 1998 - 
see also Section 4).

Models for the primordial density fluctuations have developed in response 
to our changing knowledge of the structure in the universe. The Cold Dark 
Matter (CDM) model is the most successful and long-lived of these scenarios, 
despite the fact that a suitable candidate CDM particle has yet to be 
discovered.
The amplitude of density fluctuations is constrained on large scales 
by the measurement of anisotropies in the cosmic microwave background 
radiation by the COBE satellite (Smoot etal. 1992). 
On $8 h^{-1}$Mpc scales, the local abundance of hot X-ray clusters 
provides tight constraints on the density fluctuation amplitude 
(White, Efstathiou \& Frenk 1993; Eke, Cole \& Frenk 1996).
The simplest choice of parameters in the CDM model, known as standard 
CDM, cannot simultaneously satisfy these constraints. This has led to 
interest in variants of the model such as the currently popular   
 low density universes, either with or 
without a vacuum energy or cosmological constant term.

The development of structure formation theory has been greatly aided 
by N-body simulations, which can follow the growth of density fluctuations 
well into the nonlinear regime. 
The current state of the art is represented by the 
Hubble volume simulation with one billion particles in a cube of 
$2000 h^{-1}$Mpc on a side (The Virgo Consortium - see Glanz 1998) 
and by high resolution simulations of individual dark matter 
haloes with several million particles (Moore etal. 1998a).

The conclusion reached from these studies is that there is a 
lot of evidence, though most of it circumstantial, in support of a  
hierarchical sequence of structure formation in a universe 
in which most of the mass is in the form of collisionless dark matter.

\section{Semi-analytic galaxy formation}

\begin{figure}
{\epsfxsize=13.truecm \epsfysize=12.truecm 
\epsfbox[5 240 560 600]{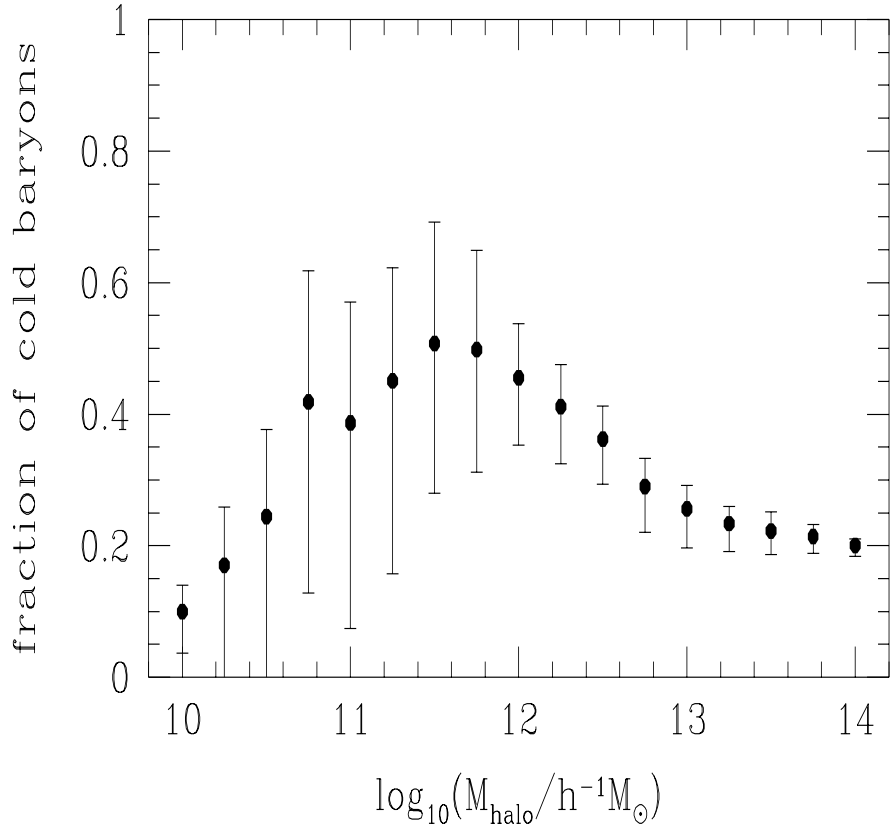}}
\caption[]
{
The fraction of baryons relative to the universal value 
in the form of cold gas or stars as a function 
of the mass of the host dark matter halo. 
The contribution of the satellite galaxies 
within a dark matter halo is included. The error bars show the 10 and 90 
percentiles. At low masses, the fraction of cold baryons is suppressed by 
feedback. High mass haloes form relatively recently, and so the gas density 
is lower and cooling less efficient.
}
\label{fig:1}
\end{figure}

The basic idea of hierarchical galaxy formation was outlined twenty 
years ago by White \& Rees (1978). Galaxies form when shock heated 
gas cools and condenses inside dark matter haloes. 
The haloes themselves form via mergers and the accretion of 
smaller sub-units.
Following on from the work of White \& Frenk (1991) and Cole (1991), 
several groups have developed semi-analytic models of galaxy formation 
that can track the entire star formation and merger histories of galaxies 
(Kauffmann, White \& Guiderdoni 1993; Lacey etal. 1993; Cole etal. 1994; 
Somerville \& Primack 1998).

There are two main components to any such model: a scheme to 
describe the growth of dark matter haloes and a set of rules to 
describe both the gas processes involved in star formation and 
the mergers of galaxies within the dark matter haloes.

The growth of dark matter haloes is the best understood of these components, 
due to the extensive tests of the analytic theories that have been made 
against the results of N-body simulations (e.g. Lacey \& Cole 1994).
The majority of models use a Monte-Carlo scheme (e.g. Kauffmann \& 
White 1993; Somerville 
\& Kolatt 1998) to generate a merger history for each dark matter halo; 
the distribution of progenitor masses 
at a given redshift can be compared with the extension of Press \& Schechter's 
(1974) theory for the mass function of bound objects, derived by Bower (1991) 
and by Bond etal. (1991). 
Alternatively, one could extract the merger history of a dark matter halo 
directly from a N-body simulation, using closely spaced output 
times (Kauffmann etal. 1998).
Statistically, the merger trees generated with a Monte-Carlo algorithm and 
those extracted from simulations appear to be equivalent; 
furthermore, the merger 
history of a dark matter halo is independent of its local environment 
(Lemson \& Kauffmann 1998).

Our present understanding of the complex physical processes 
involved in star formation can be encapsulated into a 
set of simple rules that describe the following stages: 

\begin{itemize}
\item[(i)] Gas is shock heated to the virial temperature of its host 
           dark matter halo.
\item[(ii)] Hot gas cools radiatively at a rate that depends upon its 
            density and chemical composition 
            (Rees \& Ostriker 1977; Silk 1977; 
            Binney 1977).
\item[(iii)] Cold gas turns into stars. 
\item[(iv)] Some form of regularisation or feedback of the star formation 
            process prevents all the gas from cooling in small 
            objects at high redshift, when cooling is most efficient due to 
            the high density.
\item[(v)] Stellar population synthesis gives a luminosity and colour for each 
           galaxy (Bruzual \& Charlot 1993, 1998).
\item[(vi)] Galaxies may merge together  - see Section 5.
\end{itemize}

\begin{figure}
{\epsfxsize=13.5truecm \epsfysize=17.truecm 
\epsfbox[30 0 560 750]{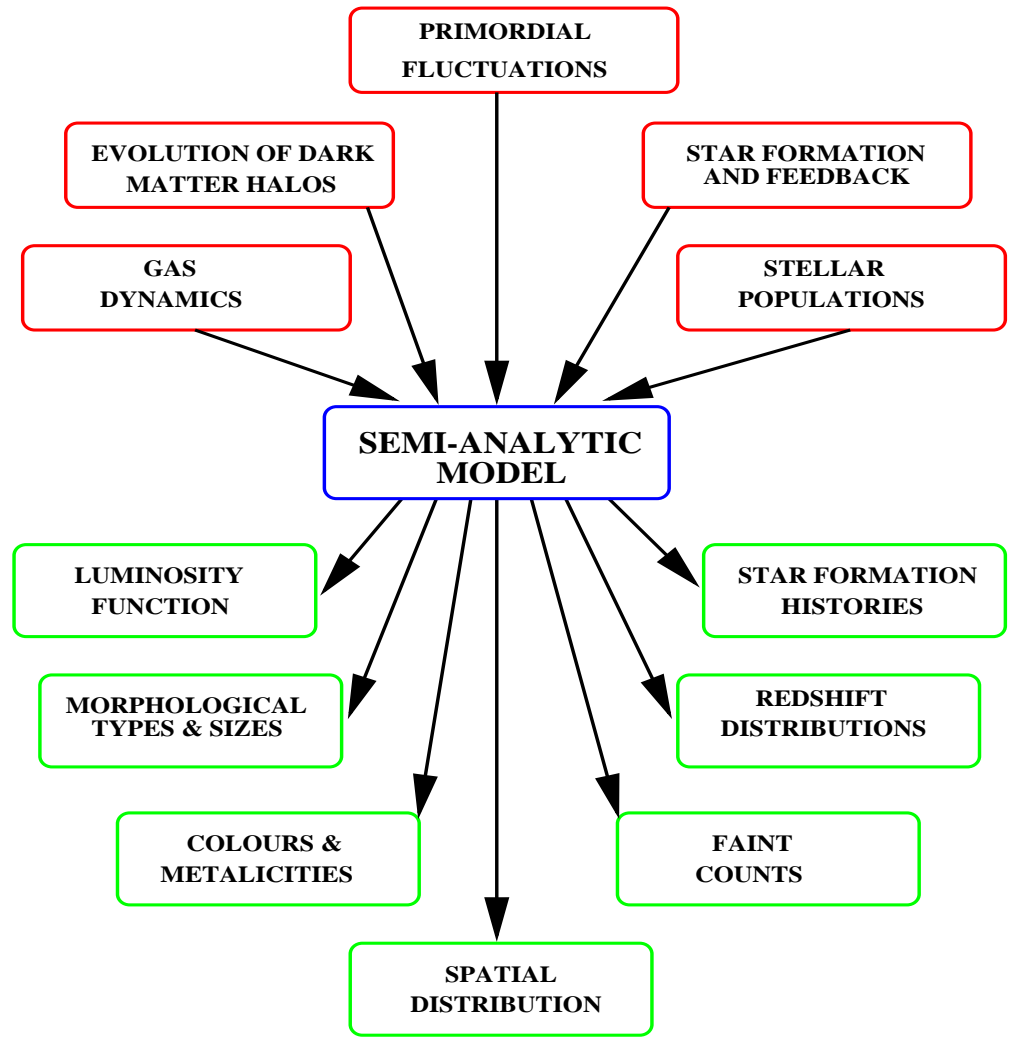}}
\caption[]
{
An overview of how the semi-analytic model works
}
\label{fig:0}
\end{figure}

The interplay of these various processes is illustrated by Figure \ref{fig:1}. 
This shows the fraction of baryons that are in the form of cold gas or stars 
as a function of the mass of the host dark matter halo. 
Within a particular halo, we have added the contribution of the 
satellite galaxies to the cold baryons in the galaxy that resides at the 
centre of the halo. 
For haloes less massive than $10^{11} h^{-1} M_{\odot}$, the fraction of 
cold baryons is low due to feedback; this is parameterised in 
such a way that star formation is suppressed more strongly in low mass 
haloes. 
Higher mass haloes form relatively recently and have a lower mean and 
central gas density; 
cooling in these haloes is therefore inefficient.

A surprisingly small number of parameters are needed to produce a fully 
specified galaxy formation model. These are set with reference to a subset of 
the observational data at low redshift; either the normalisation of the 
local luminosity function (Cole etal. 1994) or the zero-point 
of the Tully \& Fisher (1977) relation between the rotation 
speed and luminosity of spiral galaxies (Kauffmann etal. 1993). 
This gives the models tremendous predictive power. 
The Durham group have recently overhauled the model described by 
Cole etal. (1994); the main additions are a new method to generate 
merger histories for dark matter haloes, a prescription for 
chemical enrichment, a treatment of galaxy sizes and the 
inclusion of dust (Cole etal. 1998). 
The model predicts an enormous range of individual and global properties of
galaxies: bulge and disk scale lengths, bulge and disk 
magnitudes and stellar metallicities, star formation rates, morphological 
type, colour, luminosity function, number counts, redshift distributions and 
clustering.

This physical approach can be contrasted with the one adopted in 
empirical models, which extrapolate the properties of present 
day galaxies back in time.
It is instructive to count the parameters that are involved 
in such a model, for example to reproduce the number counts of faint 
galaxies.
One starts with a range of morphological types, each of which 
has a template form for the spectral energy distribution, in order to 
compute the bandshift and evolutionary corrections to the galaxy 
magnitude when viewed at different redshifts. 
Two parameters are needed to reproduce this spectrum - an initial 
time when star formation is switched on and a star formation timescale.
Next, a luminosity function is specified for each type. This requires 
three parameters, the normalisation, the characteristic luminosity and 
the faint end slope. 
If we want to use five different morphological or spectral types in 
this kind of model, we already need to specify  
$5 \times 5 = 25$ parameters. 
This is before adding an empirical dust correction or invoking 
fading populations of dwarf galaxies to improve the match with 
the observed counts. 
Furthermore, there are two important limitations of this approach 
that should be borne in mind. 
Firstly, structure formation in the universe is ignored. No account is 
taken of how much cold gas is available at high redshift, when the star 
formation rates are chosen.
Secondly, after all this effort, all that one has gained is a model for  
number counts. To predict any other galaxy properties, further ingredients 
and parameters have to be added.

\section{The star formation history of the universe}

\begin{figure}
{\epsfxsize=12.5truecm \epsfysize=15.truecm 
\epsfbox[20 170 550 720]{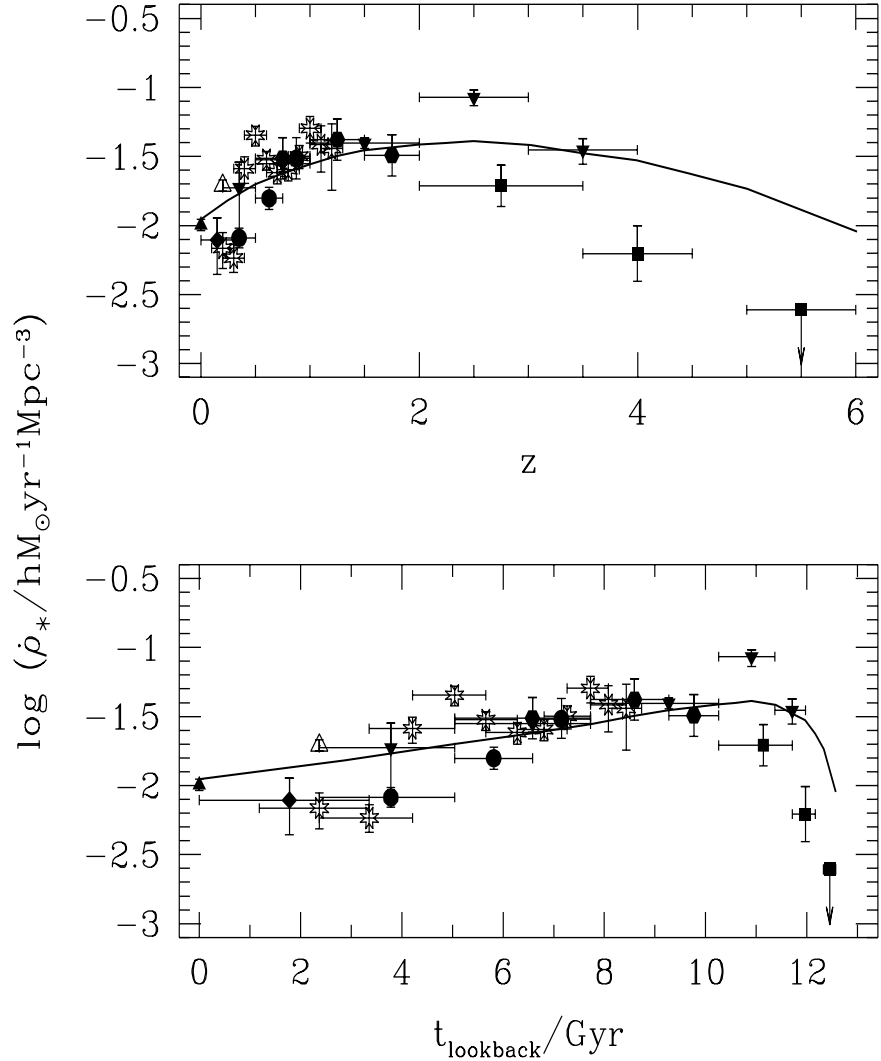}}
\caption[]
{
The global star formation history in a universe with present day 
density parameter $\Omega_{0}=0.3$ and a cosmological constant 
$\Lambda_{0}=0.7$. 
The line shows the prediction of the semi-analytic model -- the 
parameters in the model are set by comparison with the local $B$ and 
$K$ band luminosity functions.
The top panel shows the star formation rate per unit volume 
plotted against redshift. The filled symbols show a compilation of 
data -- full references are given in Baugh etal. (1998a). 
We use the IMF propsed by Kennicutt (1983) to convert from 
flux to star formation rate.
The lower panel shows the same quantity plotted as a function of 
lookback time. The age of this universe is $13.5$Gyr.
The open stars are taken from 
Hogg etal. (1998), who estimate the star formation rate from OII 
luminosity and the open triangle at $t_{lb} \sim 2.4 $Gyr is from Tresse \& 
Maddox (1998), who use the H-$\alpha$ flux density to infer 
the star formation rate.
}
\label{fig:2}
\end{figure}

Hierarchical models in which star formation in low mass haloes is suppressed 
by feedback,  naturally predict that a large fraction of the stars 
that we see today formed at fairly low redshifts.
Typically $50 \%$ of the stars form since a redshift of $z \sim 1$--$1.5$ (see 
the solid line in Figure \ref{fig:4}). 
The global star formation rate per unit volume predicted by our upgraded 
model,  in a low density universe with a cosmological constant,  
 is shown by the solid line in the panels of Figure \ref{fig:2}. 
The star formation rate is computed from the $1500\AA$ flux density and 
includes obscuration by dust; the effects of dust depend upon the 
mass of cold gas and metals in the galaxy and the angle it is viewed at.
There are small differences between this curve and that shown in 
Figure 16c of Baugh etal. (1998a), due to improvements in our 
modelling of gas cooling, the star formation timescale and feedback, 
which are described fully in Cole etal. (1998).
Note that the parameters in this model are set with reference to the 
local $B$ and $K$-band luminosity functions, and have not been 
readjusted to `fit' the data points in Figure \ref{fig:2} .
The predicted star formation rate rises with increasing redshift to 
$z=1$--$2$.
The star formation rate thereafter remains relatively flat to z$\sim 3$, 
varying by less than an order of magnitude over the entire history of the 
universe. 

New techniques for selecting high redshift galaxies from deep images 
-- the `Lyman-break' galaxies, so-called because they appear red in 
one colour, due to the redshifted Lyman-limit 
discontinuity and blue in another colour, due to ongoing star formation  
(Steidel \& Hamilton 1992; Madau etal. 1996) -- have allowed the star 
formation history of the universe to be pieced together. 
The filled symbols in the top panel of Figure \ref{fig:2} show a compilation 
of observational determinations of the global star formation rate -- full 
references are given in Baugh etal. (1998a). 

We plot the star formation history as a function of lookback time in the lower 
panel of Figure \ref{fig:2}. The area under this curve gives the mass of stars 
formed. 
The open symbols show recent determinations of the global star formation rate;
the open triangle is a measurement of the star formation rate using the 
H-$\alpha$ flux density at a lookback time of $2.4$Gyr, taken 
from Tresse \& Maddox (1998). 
The open stars are inferred from the OII luminosity density 
measured by Hogg etal. (1998).

\section{The evolution of galaxy clustering}
\begin{figure}
{\epsfxsize=14.5truecm \epsfysize=16.truecm 
\epsfbox[10 190 550 600]{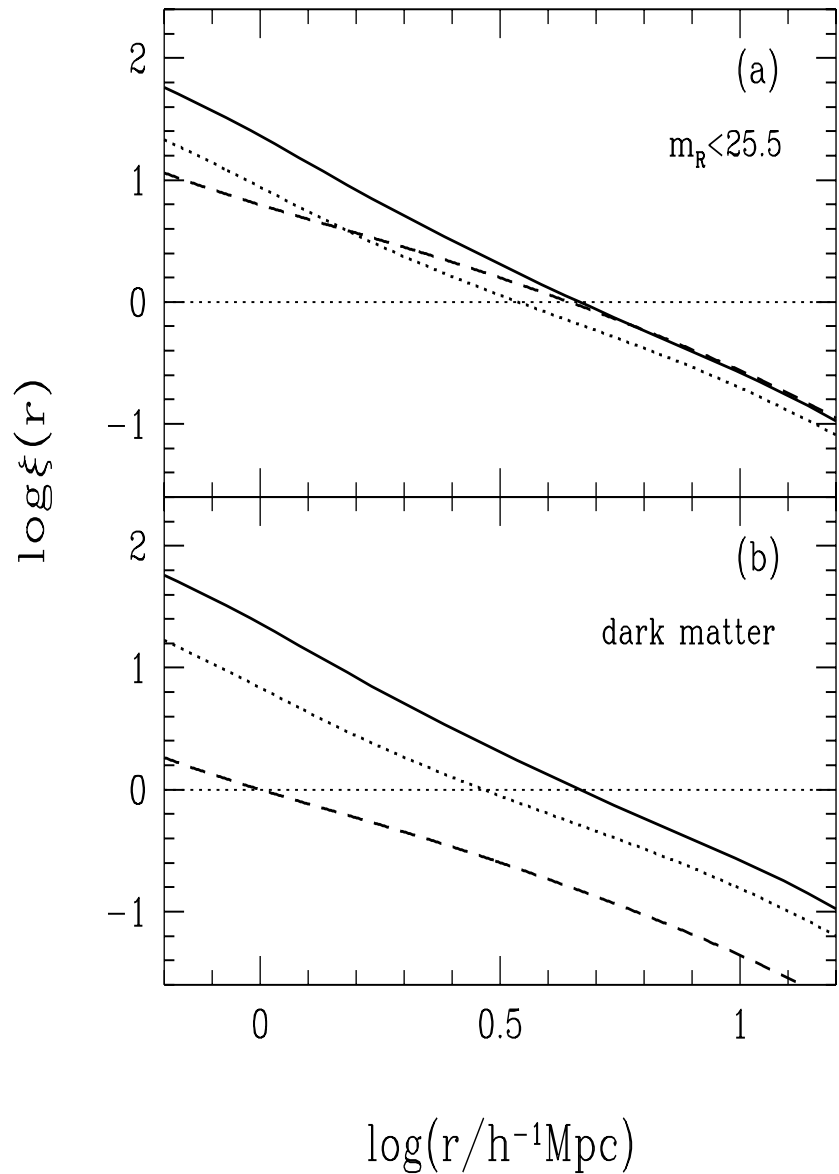}}
\caption[]
{
The correlation functions of galaxies and dark matter at 
different redshifts as a function of comoving separation: 
the solid lines correspond to  $z=0$, the dotted 
lines to $z=1$ and the dashed lines to $z=3$.
(a) The correlation function of galaxies brighter than an 
apparent $R$-band magnitude of $m_{R} = 25.5$. 
(b) The correlation function of the underlying dark matter. 
The horizontal dotted line is shows $\xi(r_{0})=1$, which 
defines the correlation length, $r_{0}$.
}
\label{fig:3}
\end{figure}

We can apply exactly the same colour selection technique to our model 
galaxies that is applied to deep images of the sky to isolate high 
redshift galaxies.
Our model predicts that galaxies that can be detected as Lyman-break 
galaxies reside in dark matter haloes of mass $\sim 10^{12}h^{-1}M_{\odot}$. 
We can compute the correlation function of Lyman break 
galaxies in our model (Baugh etal. 1998a), 
using a simple analytic prescription 
for the bias of their host dark matter haloes (Mo \& White 1996; see 
also Cole \& Kaiser 1989). 
The Lyman break galaxies are predicted to have a correlation length, in 
comoving units, of around $4h^{-1}$Mpc, similar to that of bright galaxies 
at the present day (e.g. Baugh 1996).
However, at $z\sim 3$, such a correlation length implies strong clustering 
relative to the dark matter; we find that the bias parameter of 
Lyman break galaxies in a critical density universe is 
$b \approx 4$ -- this means that the amplitude of the correlation 
function of Lyman-break galaxies is more than an order of magnitude 
greater than that of the dark matter.
Governato etal. (1998) combined our semi-analytic model for galaxy formation 
with an N-body simulation to obtain the spatial distribution of Lyman break 
galaxies. This allowed the clustering of the Lyman break galaxies to 
be measured directly; the results are in good agreement with the simple 
analytic calculation carried out by Baugh etal. (1998a).
The predicted correlation length was confirmed by the 
subsequent measurements of the clustering of Lyman break galaxies 
(Steidel etal. 1998; Giavalisco etal. 1998; Adelberger etal. 1998).

In general, the correlation function of galaxies evolves with redshift in 
a much more complicated fashion than the correlation function of the 
dark matter. 
The amount of clustering that is measured is sensitive to the 
observational selection applied to the galaxy population. 
At a fixed apparent magnitude limit, galaxies at higher redshift have higher 
intrinsic luminosities, and so are predicted by our 
model to reside in  progressively rarer dark matter haloes, 
compared with the general population of galaxies 
in place at that redshift. This means that the high redshift galaxies 
tend to be biased tracers of the dark matter distribution i.e. they 
exhibit stronger clustering than the dark matter 
because they form at the rare peaks of the density field.
We contrast the evolution of the galaxy correlation function with that 
of the dark matter in Figure \ref{fig:3} (Baugh etal. 1998b). 
The solid lines correspond to $z=0$, the dotted lines to $z=1$ and the 
dashed lines to $z=3$. The comoving correlation length of the dark matter, 
defined as the length scale at which the correlation function is equal to 
unity, changes by a factor of $\sim 5$ over this redshift interval.
The change in the galaxy correlation function is much smaller; the galaxy 
correlation length decreases slightly by $z=1$ before rising to attain 
essentially the  present day value at $z=3$. 
The exact factor by which the correlation 
length changes is sensitive to the magnitude limit applied.

\section{The formation of early type galaxies}

A natural model for the formation of elliptical galaxies in a 
universe in which structure formation is hierarchical is through 
galaxy mergers.
This is sometimes called the nurture hypothesis -- the environment 
of a galaxy rather than initial conditions determines its final morphology.

There are a number of pieces of observational evidence in support 
of the nurture hypothesis. 
Kauffmann, Charlot \& White (1996) used data from the Canada-France 
Redshift Survey (Lilly etal. 1995) and from the Hawaii Deep Survey 
(Cowie etal. 1996)  to study the fraction of galaxies that have the 
colour of passively evolving ellipticals out to $z\sim1$.
The fraction of galaxies with these colours falls to one third 
of the present day value by $z \sim 1$, indicating either that 
the present day ellipticals 
have not all been assembled by $z=1$, or that there is some ongoing star 
formation that changes their colour from that expected for an old, passively 
evolving population.
Zepf (1997) has analysed deep optical and infrared images to search for 
very red galaxies. If ellipticals formed all their stars in a burst 
at high redshift, they would have rapidly attained a red colour 
after the burst 
epoch. In this case, given the local abundance of early type galaxies, 
there should be many times more red objects per square 
degree on the sky than are observed.

The formation of elliptical galaxies in semi-analytic models is 
outlined by Kauffmann, White \& Guiderdoni (1993) 
and Baugh, Cole \& Frenk (1996a). 
The models assume that gas initially cools into a disk and forms 
stars. 
When two dark matter haloes merge, the galaxies within them can 
coalesce on a quite different timescale. 
If the dynamical friction timescale for a galaxy is shorter than the 
lifetime of the dark matter halo it resides in (the lifetime is usually 
defined as the time until the mass of the halo doubles through accretion 
and mergers), then the galaxy merges with the central galaxy in the 
halo. 
Numerical simulations indicate that some fraction of the accreted 
satellite mass moves to the central region of the galaxy and the rest 
is spread out over the disk of the central galaxy (Walker etal. 1996).
In our model, we assume that all the accreted stellar material is 
transferred to the bulge of the central galaxy and that the cold gas of the 
satellite is added to the disk.
If the accreted satellite is above a certain fraction of the mass in 
cold baryons of the central galaxy, typically taken to be in the 
range 30\%--50\%, then the merger is termed  a violent merger. 
In this case, the disk of the central galaxy is destroyed and 
the stars are moved to the bulge. Furthermore, any cold gas present 
is turned into stars in a burst and added to the bulge. 
Hence, immediately after a violent merger, the galaxy will have a pure 
bulge morphology. 
Subsequently, more cooling gas can be accreted in a disk and turned into 
stars, reducing the bulge to disk ratio of the galaxy; this is 
the quantity that we use to assign a morphological type to the galaxy. 
The morphology of a galaxy is therefore a function of time in our model.

\section{The predicted properties of early type galaxies}

\begin{figure}
{\epsfxsize=12.5truecm \epsfysize=12.truecm 
\epsfbox[0 200 550 680]{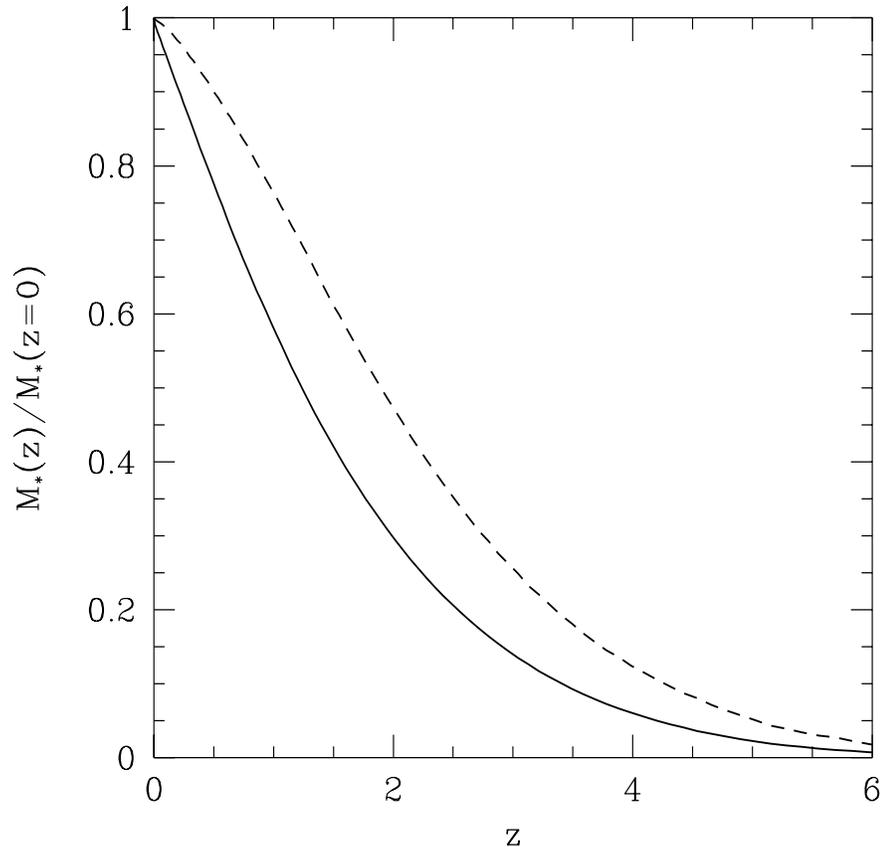}}
\caption[]
{
The fraction of present day stars already place 
by a given redshift.
The solid line shows the build up of stellar mass in the 
field i.e. counting galaxies in dark matter haloes of all masses. 
The dashed line shows the build up of stellar mass in only those 
galaxies that reside in haloes more massive 
than $ 5 \times 10^{14} h^{-1} M_{\odot}$, corresponding to clusters.}
\label{fig:4}
\end{figure}

\subsection{The epoch of bulge formation}
 
The epoch of bulge formation may be defined as the formation time 
of the stars that make up the bulge or the time at which the bulge 
is actually assembled -- these times can be very different 
(see Figure 3 of Kauffmann 1996).

Star formation in galaxies that reside in a cluster at the present 
day is advanced with respect to the field. This is shown by the dashed 
line in Figure \ref{fig:4}, which shows the fraction of the present day 
stellar mass in place by a given redshift, for galaxies found today in 
haloes with velocity dispersions in excess of a $1000 {\rm kms}^{-1}$. 
Half the stars are in place by a redshift of $ z \sim 2$, compared with 
a redshift of just over unity for field galaxies.
The collapse of the haloes in which the cluster galaxies form is
shifted to higher redshifts because they are part of an unusually large 
overdensity  -- the cluster halo -- today. 

The epoch of the last major merger depends upon the present day 
environment of the galaxy. 
The median redshift of the last violent or major merger event 
is $z_{med} \sim 0.7$ for field ellipticals in a critical density universe. 
For ellipticals found in clusters at the present day, the median redshift 
of the last major merger is higher, $z_{med} \sim 0.9$ (Baugh etal. 1996a). 
The amount of cold gas involved in a merger falls as the redshift of the 
merger decreases. This is a result of the global cold gas fraction 
peaking around a redshift of $z=1$--$2$, the epoch when more 
massive haloes start to collapse in significant numbers -- the 
feedback associated with star formation in these haloes is less 
severe than it is in low mass systems (see Figure 17 of Baugh etal. 1998a); 
the cold gas fraction declines after $z \sim 1$ when several e-folding 
times of the star formation timescale have elapsed.
Typically only $5\%$ of the final stellar mass of an elliptical is formed 
in the burst that accompanies a major merger,  
if the last major merger takes place after $z \sim 0.5$; between $z=0.5$ and 
$z=1.0$, the burst of star formation that accompanies a major merger 
can account for $15\%$--$20\%$ of the final stellar mass.
The stars formed in the burst could be responsible for the 
`intermediate age populations' detected in some ellipticals (e.g. 
Rose \& Tripicco 1986).

The picture that emerges from the semi-analytic models is that 
the stellar content of the spheroidal component of a galaxy is 
formed at high redshift, in smaller fragments. The spheroid is 
assembled in a violent merger at relatively low redshift.

\subsection{The colour magnitude relation of cluster ellipticals}

Early-type galaxies in clusters display a remarkably tight 
sequence of colour as a function of intrinsic brightness 
(Bower, Lucey \& Ellis 1992; Gladders etal. 1998 - see also 
the contributions of Richard Bower and Omar Lopez-Cruz in this 
volume).
This has been interpreted as a strong indication for a coeval 
formation epoch for the ellipticals and places strong constraints 
on any recent star formation.

The semi-analytic models produce a similarly small scatter in 
the colour-magnitude relation (Baugh etal. 1996a; Kauffmann 1996). 
The explanation for this is that although the assembly of the 
bulge or elliptical may be a relatively recent event, the stars 
involved are old. 
Baugh etal. obtained a flat colour magnitude relation -- this 
version of the semi-analytic model did not include chemical 
evolution and only used a solar metallity stellar population model 
(Bruzual \& Charlot 1993).
Kauffmann \& Charlot (1998) demonstrated that the observed slope of 
the colour-magnitude relation can be recovered once chemical enrichment 
is accounted for. These authors argue that big elliptical galaxies are 
produced by the merger of big spiral galaxies -- galaxies in deeper 
potential wells are better at retaining the metals produced in star 
formation. The tilt of the colour-magnitude relation is 
thus interpreted as a result of a sequence in metallicity with 
intrinsic luminosity.

The scatter in the colour-magnitude relation recovered in the 
models changes little out to a redshift of $z \sim 0.5$ 
(Baugh etal. 1996a). There is a general blueing of the galaxies, 
which matches that observed (Arag\'{o}n-Salamanca etal. 1993).

\subsection{The evolution of cluster membership}

Clusters at high redshift are observed to have a higher fraction 
of blue or spiral galaxies than low redshift clusters (e.g. 
Butcher \& Oemler 1984; Dressler etal. 1994). 
One important point to be wary of when comparing observations of clusters 
at different redshifts, is that an object of a given mass today  
would correspond to a much {\it rarer} object if it was found at 
higher redshift.
It is necessary to take into account the effect this would have on the rate 
at which the cluster mass is assembled and on the star formation in the 
progenitor haloes.

Many mechanisms have been proposed for changing the morphologies of 
galaxies once they fall inside a cluster potential e.g. the  
harassment of spiral galaxies through numerous impulsive encounters 
with other galaxies in the cluster (Moore etal. 1998b).
An alternative explanation is that the distribution of morphological 
types of the galaxies that become cluster members changes with 
redshift (Kauffmann 1995; Baugh etal. 1996a). 
The number of elliptical galaxies 
that are in place to become cluster members at $z=0.5$ is lower 
than that at $z=0$. This results from an interplay among 
the timescales for gas cooling, star formation and galaxy mergers; 
by $z=0$ more cold baryons are in place and there is more time 
for galaxy mergers to have occured.

\subsection{Number counts of early-type galaxies}

The one-tenth arcsecond resolution of images from the Hubble Space 
Telescope has made it possible to divide the number counts of faint 
galaxies into the contributions from different morphological types 
(e.g. Driver etal. 1995, Glazebrook etal. 1995). 
The counts obtained from our semi-analytic model are in remarkably good 
agreement with the observations, especially in view of the fact that 
the one free parameter involved in determining the morphological types -- 
the threshold used in the definition of a violent merger -- is set so that 
the model reproduces the local morphological mix (Baugh etal. 1996b).

\section{Summary}
We have argued in this review that our knowledge of structure 
formation has improved tremendously in the past twenty years, 
and that it would seem both prudent and necessary to incorporate 
the growth of structure when building models for galaxy formation.
Several groups around the world have developed semi-analytic 
prescriptions for galaxy formation with this in mind. 
The aim of these models is ambitious -- to follow the entire star formation 
and merger history of a galaxy, from the collapse of primordial 
density fluctuations to the present day. In such models, the 
`identity' of a galaxy is not indelible -- a luminous galaxy today 
could have been in several fragments at $z=1$ or it may have been in 
one, much smaller piece. 

The power of the semi-analytic technique is illustrated by the example 
provided by Lyman break galaxies. We are able to take the same filters 
used by the observers and apply them to the galaxies in our model. 
Real galaxies and model galaxies are subject to the same selection criteria.
A range of models recover the observed abundance of Lyman break galaxies, 
which is very sensitive to the choice of stellar IMF or to the normalisation 
of the density fluctuations. The models predict the size of the Lyman-break 
galaxy, the stellar mass, star formation rate and the mass of 
the host dark matter halo. 
The halo mass is relatively large compared to the rest of the dark matter 
halo population in place by this redshift -- this led to the prediction 
that the Lyman break galaxies should be strongly clustered relative to the 
dark matter, which was confirmed by subsequent observations.

One of the themes running through this meeting was that the classical 
or monolithic picture and the `merger' model of elliptical galaxy 
formation in clusters give essentially the same result -- in both cases 
the stars that make up an elliptical galaxy are old; the main 
difference between the two models is the time when the stars are actually 
arranged into their final morphology.
This represents a triumph for the hierarchical galaxy formation model. 
The classical model is designed to explain one observation -- 
the small scatter in the colour-magnitude relation for cluster ellipticals.
This is not a difficult result to achieve in isolation.
The parameters in the semi-analytic model described here were set to 
reproduce as closely as possible the local field luminosity functions. 
The colour magnitude relation is simply one of the many outputs 
that come out of the model. 
This would seem to vindicate the physically motivated approach we 
have taken and suggests that many of the key ingredients 
of a successful theory of galaxy formation are in place, though 
the details still have to be worked out.

\acknowledgments
CMB acknowledges financial assistance from the organisers to make 
attendance at this meeting possible.

\end{document}